\documentclass[journal]{IEEEtran}
\usepackage{amsmath,amsfonts}
\usepackage{algorithmic}
\usepackage{algorithm}
\usepackage{array}
\usepackage[caption=false,font=normalsize,labelfont=sf,textfont=sf]{subfig}
\usepackage{textcomp}
\usepackage{stfloats}
\usepackage{url}
\usepackage{verbatim}
\usepackage{graphicx}
\usepackage{cite}
\usepackage{color}
\usepackage{soul}
\usepackage{hyperref}

\begin{document}

\title{MVD: A Multi-Lingual Software Vulnerability Detection Framework}

\author{\IEEEauthorblockN{Boyu Zhang\IEEEauthorrefmark{1}\thanks{\IEEEauthorrefmark{1}The work was conducted when Boyu Zhang was a postdoctoral researcher at CREST and the University of Adelaide, Australia.},
Triet Huynh Minh Le\IEEEauthorrefmark{2}, and
M. Ali Babar\IEEEauthorrefmark{2}\IEEEauthorrefmark{3}}

\IEEEauthorblockA{\IEEEauthorrefmark{1}TikTok, Australia}

\IEEEauthorblockA{\IEEEauthorrefmark{2}CREST - Centre for Research on Engineering Software Technologies and The University of Adelaide, Australia}
\IEEEauthorblockA{\IEEEauthorrefmark{3}Cyber Security Cooperative Research Centre, Australia}

\IEEEauthorblockA{boyu.zhang68@gmail.com, \{triet.h.le, ali.babar\}@adelaide.edu.au}
}

\markboth{Journal of \LaTeX\ Class Files,~Vol.~14, No.~8, August~2021}
{Shell \MakeLowercase{\textit{et al.}}: A Sample Article Using IEEEtran.cls for IEEE Journals}

\maketitle

\begin{abstract}
Software vulnerabilities can result in catastrophic cyberattacks that increasingly threaten business operations. Consequently, ensuring the safety of software systems has become a paramount concern for both private and public sectors. Recent literature has witnessed increasing exploration of learning-based approaches for software vulnerability detection. However, a key limitation of these techniques is their primary focus on a single programming language, such as C/C++, which poses constraints considering the polyglot nature of modern software projects. Further, there appears to be an oversight in harnessing the synergies of vulnerability knowledge across varied languages, potentially underutilizing the full capabilities of these methods. To address the aforementioned issues, we introduce MVD -- an innovative multi-lingual vulnerability detection framework. This framework acquires the ability to detect vulnerabilities across multiple languages by concurrently learning from vulnerability data of various languages, which are curated by our specialized pipeline. We also incorporate incremental learning to enable the detection capability of MVD to be extended to new languages, thus augmenting its practical utility. Extensive experiments on our curated dataset of more than 11K real-world multi-lingual vulnerabilities substantiate that our framework significantly surpasses state-of-the-art methods in multi-lingual vulnerability detection by 83.7\% to 193.6\% in PR-AUC. The results also demonstrate that MVD detects vulnerabilities well for new languages without compromising the detection performance of previously trained languages, even when training data for the older languages is unavailable.
Overall, our findings motivate and pave the way for the prediction of multi-lingual vulnerabilities in modern software systems.
\end{abstract}


\section{Introduction}
\IEEEPARstart{S}{oftware} vulnerabilities refer to weaknesses or flaws in a software system that, if exploited, can compromise the system's security and functionality \cite{harzevili2023survey}. For instance, the infamous Heartbleed vulnerability in OpenSSL allowed unauthorized users to access sensitive data, leading to widespread security breaches and data theft across the internet \cite{durumeric2014matter}. With cyberattacks emerging as primary contributors to revenue loss for many businesses \cite{anderson2013measuring}, safeguarding software systems has become a paramount challenge in practice.

To address the challenges posed by software vulnerabilities, both program analysis-based and learning-based methods have been introduced.
Conventional program analysis tools analyze source code using predefined rules and patterns to detect vulnerabilities (e.g.,~\cite{bishop2007penetration,godefroid2012sage,bessey2010few}).
On the other hand, recent learning-based approaches (e.g.,~\cite{chakraborty2021deep,li2021sysevr,li2018vuldeepecker,russell2018automated,fu2022linevul,nguyen2022regvd}), with a particular emphasis on Deep Learning paradigms, have garnered significant attention for their improved efficacy over program analysis counterparts. These learning-based techniques predominantly operate by fine-tuning neural network models, typically pre-trained language models, using established supervised datasets tailored for vulnerability detection. The primary objective of this process is to minimize the discrepancy between model's predictions and ground-truth labels.

While most learning-based techniques have been developed and demonstrated effective for the C and C++ languages, there is little study on the performance of these models when applied to detecting vulnerabilities in other languages.
In fact, many of the real-world projects are not written in C/C++, thus limiting the direct usage of the current models for vulnerability prediction.
In addition, there is an increasing number of applications that are written in multiple languages, namely polyglot projects~\cite{mussbacher2024polyglot}.
It is also worth noting that such projects not (only) written in C/C++ still have serious vulnerabilities that potentially lead to catastrophic consequences~\cite{livshits2005finding,li2022vulnerability,alfadel2023empirical}.
As a result, current models predicting vulnerabilities in a single language like C/C++ would have limited applications in modern software development environments.

To address the aforementioned challenges, we introduce an innovative \underline{\textbf{M}}ulti-lingual \underline{\textbf{V}}ulnerability \underline{\textbf{D}}etection (\textbf{MVD}) framework. Unlike conventional models confined to a specific programming language, our framework is uniquely designed to detect software vulnerabilities across multiple languages, making it particularly suitable for contemporary software projects that often involve several programming languages.
MVD is designed based on the observations that many types of vulnerabilities, such as buffer overflow and SQL injection, manifest across a multitude of languages.
By harnessing this ubiquitous vulnerability knowledge, our framework can assimilate and transfer knowledge across diverse languages, thereby fostering a more holistic understanding of vulnerabilities and enhancing detection performance.

Under the MVD framework, we construct a multi-class classifier to discern not only the vulnerability of the input source code but also the specific language of the vulnerability. The auxiliary task of language classification augments vulnerability detection by enhancing the model's contextual understanding of language-specific vulnerability patterns. The classifier leverages CodeBERT~\cite{feng2020codebert} to extract syntactic and semantic features from multi-lingual source code. Given the disparity in the volume of labeled vulnerable data across languages, we also augment the model with a specialized loss function to address this class imbalance issue.
Furthermore, we incorporate an incremental learning module into MVD to allow it to adapt seamlessly to new languages on which it was not initially trained.
Our exhaustive experiments and ablation studies show that our model significantly surpasses single-language vulnerability detection baselines, attesting to the efficacy of our proposed model.

Our key \textbf{contributions} can be summarized as follows:

\begin{itemize}

\item To the best of our knowledge, we present MVD -- the first Deep Learning based framework for multi-lingual vulnerability detection, which has been under-explored in the current literature.

\item We have extensively evaluated MVD on 11K+ real-world vulnerabilities in six programming languages, namely Python, Java, C/C++, C\#, TypeScript, and JavaScript. We demonstrate that MVD outperforms the (single-language) state-of-the-art vulnerability prediction models by 83.7\% -- 193.6\% in terms of PR-AUC across the six prominent programming languages, with all the proposed components contributing meaningfully to the model's overall performance. Our MVD model is also effectively and efficiently extensible to new languages, even outperforming the single-language counterparts in four out of six languages.\footnote{We use the term \textit{MVD model} to refer to the multi-lingual vulnerability prediction model in the MVD framework.}
Remarkably, this extension mostly has modest to no impact on previously trained languages, even without their original training data.

\item We make our data, models, and code publicly available to support future research in multi-lingual vulnerability prediction at~\url{https://figshare.com/s/10ec70108294a225f391}.
\end{itemize}

\noindent \textbf{Paper structure}. Section~\mbox{\ref{sec:background}} introduces software vulnerability (SV) detection and the missing consideration of multi-lingual SV detection. Section~\mbox{\ref{sec:mvp}} presents the proposed MVD model for multi-lingual SV detection.
Section~\mbox{\ref{sec:setup}} describes the settings of empirical evaluation of MVD.
Section~\mbox{\ref{sec:results}} reports the experimental results of performance evaluation of MVD in different settings.
Section~\mbox{\ref{sec:threats_to_validity}} discusses the threats to validity. Section~\mbox{\ref{sec:conclusions}} concludes the study.

\section{Background, Related Work, and Motivation}\label{sec:background}

\subsection{Learning-based Vulnerability Prediction}
\label{subsec:function_svp}

In recent years, learning-based approaches have been widely used to automate the identification/prediction of SVs in source code~\cite{lin2020software,hanif2021rise}. The predictions have been performed on various levels of granularity, ranging from package/file to function and line. Among these levels of granularity, the function level is the most investigated as it reduces inspection effort for developers while still providing sufficient context of code for prediction~\cite{hin2022linevd,fu2022linevul}. Thus, the function-level prediction is also adopted for our multi-lingual investigations.

Deep Learning has been increasingly investigated for function-level vulnerability prediction~\cite{zeng2020software}. Recurrent Neural Networks like Long-Short Term Memory have been initially used for the task because of their ability to capture long-term dependencies in code (e.g.,~\cite{li2018vuldeepecker,russell2018automated,li2021sysevr}). Later, to more precisely capture the structure and semantic meaning of code, graph-based models, including Gated Graph Neural Networks employed in Devign~\cite{zhou2019devign}, ReVeal~\cite{chakraborty2021deep} or Graph Convolutional Networks used in IVDetect~\cite{li2021vulnerability}, Graph Attention Network in LineVD~\cite{hin2022linevd}, have been explored for function-level vulnerability prediction. These graph-based models have shown superior performance than LSTM.
Recently, LineVul~\cite{fu2022linevul}  relying on CodeBERT~\cite{feng2020codebert}, a pre-trained large language model, has demonstrated the state-of-the-art performance for function-level vulnerability prediction~\cite{steenhoek2023empirical}, outperforming various recurrent and graph-based neural networks. It is important to note that LineVul has only been evaluated on C/C++ vulnerabilities, and thus, its ability to perform multi-lingual vulnerability prediction is still largely unknown.

\subsection{Missing Consideration of Multi-Lingual Vulnerability Prediction}

As mentioned in Section~\ref{subsec:function_svp}, function-level vulnerability prediction has gained significant traction in the recent literature, but the latest advances, particularly using Deep Learning, for this task have mostly focused on detecting vulnerabilities in C/C++.
However, we argue that multi-lingual vulnerability prediction is crucial in modern software development because of the following three reasons.
Firstly, many large and widely used software systems are not written only in C/C++. For example, many mobile apps are written in Java; modern web development mostly requires JavaScript and TypeScript; Artificial Intelligence-based systems are frequently written in Python; the development of video games heavily relies on C\#.
Secondly, contemporary software projects are increasingly complex, often incorporating multiple programming languages to leverage the strengths of each, a.k.a. polyglot projects~\cite{mussbacher2024polyglot}. Specifically, Mayer et al.~\cite{mayer2015empirical} found that polyglot projects are prevalent in practice, averaging five languages per project. This finding was later confirmed in a follow-up study through a survey with 139 software professionals~\cite{mayer2017multi}.
Thirdly, the most dangerous vulnerabilities according to the top-25 CWE-IDs list in 2023,\footnote{\url{https://cwe.mitre.org/top25/archive/2023/2023_top25_list.html}} are mostly language agnostic, except for NULL Pointer Dereference (CWE-476) only applicable to languages utilizing pointers like C/C++ and Code injection (CWE-94) only applying to interpreted languages. This means that languages other than C/C++ can also be subjected to high-impact vulnerabilities like the Log4Shell vulnerability~\cite{log4j_vuln}  in Java recently.
All of the three aforementioned observations show a dire need for approaches that can perform vulnerability prediction in multiple languages.

Despite the aforementioned benefits, multi-lingual vulnerability prediction poses three key challenges to be addressed. Firstly, the number of programming languages is large, so training and maintaining a separate model for each language as per the current practice is quite resource-intensive and inefficient in practice. A more practical approach is to develop a single model that can consume data in different languages and predict new vulnerabilities in respective languages. Nevertheless, the effectiveness of such a combined model has not been investigated.
Secondly, different languages have distinct code syntax, creating potential issues for code representation. An effective representation model needs to capture the nuances in various languages without requiring significant changes to the model architecture.
Code models adapted from large pre-trained language models like CodeBERT~\cite{feng2020codebert} are advantageous in this scenario because they have the demonstrated ability to capture syntactic and semantic information of code in different languages through masked language modeling~\cite{xu2022systematic}.
However, the effectiveness of large language/code models for multi-lingual vulnerability prediction has not been well understood.
Thirdly, new languages emerge over time, making it expensive to frequently retrain a model from scratch for the new languages. A better way would be to reuse the knowledge of a trained model on existing languages to adapt to a new language. In this case, we only need to train the model on the data of the new language, which would significantly reduce the training time. This process is commonly referred to as \textit{incremental learning} or \textit{continual learning}~\cite{de2021continual}. However, the use of incremental learning to handle 
new languages for multi-lingual vulnerability prediction is yet to be explored.
Overall, to the best of our knowledge, our study is the first to address the above three challenges, aiming to propose an effective and efficient solution to multi-lingual vulnerability prediction.

\subsection{Incremental Learning in Software Engineering}
Incremental learning has become an increasingly important yet under-explored area of research within Software Engineering, particularly for tasks that involve evolving datasets and the need for models to adapt over time without forgetting previous knowledge. This concept is crucial in software engineering due to the dynamic nature of software development and the continuous integration of new code and features.

Pamela et al. \cite{bhattacharya2010fine} presented an innovative approach that combines incremental learning with multi-feature tossing graphs. This method allows for the continuous updating of the bug triage system with new data, improving its accuracy and efficiency over time. By incorporating fine-grained incremental learning, the system can adapt to new patterns in bug reports and developer activities without discarding the valuable knowledge accumulated from historical data. Zi et al. \cite{yuan2013predicting} applied incremental learning to the prediction of bugs in source code changes. This approach is particularly relevant in continuous integration and deployment environments, where code changes are frequent and models must rapidly adapt to new data. By employing incremental learning, the model can update its predictions based on the most recent changes, maintaining high accuracy in bug prediction over time. More recently, Jingmei et al. \cite{li2020incremental} addressed the challenge of classifying malware in scenarios where only limited data is available. This work leverages incremental learning to effectively update the classification model as new malware samples are discovered, ensuring that the model remains current and effective without the need to be retrained from scratch on the entire dataset. 

These prior studies have demonstrated the potential of incremental learning in addressing the challenges of software engineering tasks that require continuous adaptation. Our work adds to the body of knowledge by investigating the capability of incremental learning for vulnerability prediction. Specifically, to the best of our knowledge, we are the first to leverage incremental learning to enable a multi-lingual vulnerability prediction model to handle a new language without resource-intensively retraining the model on the data of existing languages.

\section{MVD: A Framework for Multi-Lingual Software Vulnerability Detection}
\label{sec:mvp}
In this section, we present the MVD approach for multi-lingual SV detection and elucidate its adaptability to incorporate new languages.

\subsection{Overview}

\begin{figure*}[t]
\centering
\includegraphics[scale = 0.6]{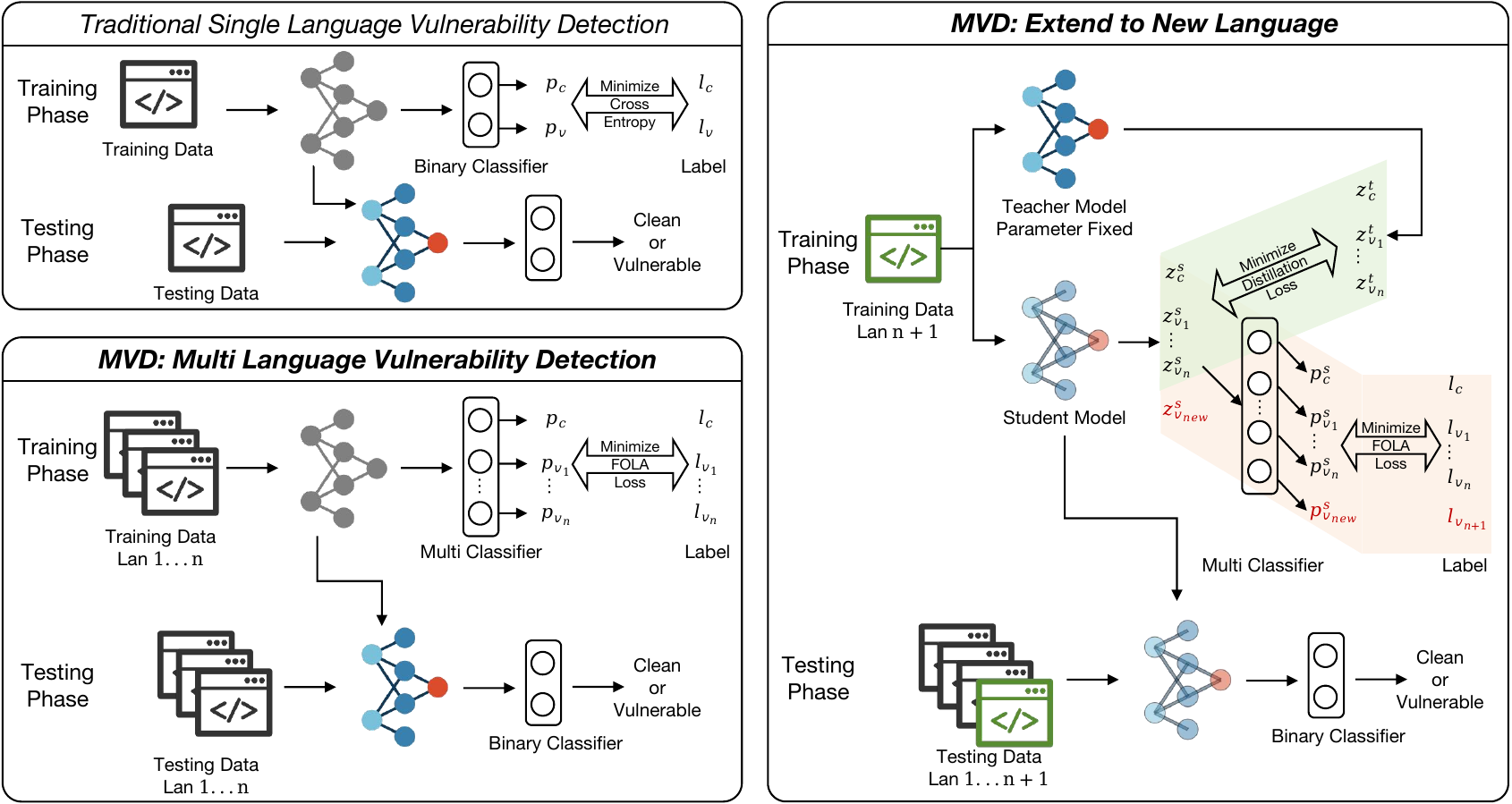}
\caption{An overview architecture of our MVD framework for multi-lingual vulnerability prediction as compared to the traditional approach for single-language vulnerability prediction.}
\label{fig:framework}
\end{figure*}

MVD aims at detecting software vulnerabilities in multiple programming languages \textit{simultaneously}. Specifically, our model is designed to detect function-level vulnerabilities in different languages. In contrast, as illustrated in Fig. \ref{fig:framework}, conventional vulnerability detection models are often constrained by their language-specific designs, where each model is typically tailored for a single programming language. This results in the need for separate models for each new language, introducing redundancy and leading to inconsistent vulnerability detection mechanisms across different languages.

To enable multi-lingual vulnerability prediction, MVD capitalizes on the CodeBERT~\cite{wu2016google} pre-trained language model. CodeBERT was chosen because it has been pre-trained with CodeSearchNet~\cite{husain2019codesearchnet}, a large code corpus of multiple programming languages, which allows CodeBERT to discern and encapsulate the intricate lexical and logical nuances of diverse code snippets, yielding a detailed vector representation.
It is also worth noting that CodeBERT is currently the state-of-the-art model for function-level vulnerability prediction~\cite{steenhoek2023empirical}.
However, MVD is unique and innovative in two significant ways compared to existing vulnerability prediction models using CodeBERT (e.g.,~\cite{fu2022linevul,hin2022linevd}). 

Firstly, MVD's training phase involves processing labeled vulnerability datasets from multiple languages simultaneously, adopting a multi-class classification approach. During training, MVD is trained to differentiate multiple classes corresponding to the existence of vulnerabilities in different languages. Note that non-vulnerable/clean code is a separate class. For inference, these vulnerable classes are consolidated into a single "vulnerable" category, enabling the model to perform a binary classification task. This approach allows MVD to assimilate shared vulnerability patterns across languages, enhancing its detection performance.

Secondly, the utilization of incremental learning in the MVD framework ensures it can effortlessly accommodate new languages without extensive retraining. This positions MVD as a progressive solution in software vulnerability detection, primed to adapt to the ever-evolving programming language landscape.

\subsection{Tokenization}
\label{subsec:tokenization}
We use the WordPiece \cite{wu2016google} tokenizer, which is aligned with the CodeBERT pre-trained model employed in our framework. WordPiece is a data-driven tokenization method that iteratively breaks down words into commonly occurring subwords or merges frequent subwords. While it shares similarities with the Byte-Pair Encoding (BPE) \cite{sennrich-etal-2016-neural} method, where frequent pairs of characters are merged into single tokens, there are subtle differences. Specifically, while BPE focuses on merging the most frequent character pairs, WordPiece prioritizes subwords based on the likelihood of their occurrence in the data. This distinction allows WordPiece to be more adaptive in representing rare words. In the context of source code vulnerability detection, WordPiece's ability to handle out-of-vocabulary words by representing them as a sequence of subwords is invaluable. It ensures that even unique identifiers and terminologies in source code can be meaningfully represented, bolstering the model's accuracy in vulnerability detection.

\subsection{Model Architecture}
Our MVD model inherits the initial weights from the pre-trained CodeBERT~\cite{feng2020codebert}, which provides a robust foundation for understanding programming languages. Upon receiving source code, the model begins by tokenizing the input using the WordPiece tokenizer (see Section~\ref{subsec:tokenization}). The tokenized results are then passed through a word embedding layer, which maps each token to a high-dimensional space, capturing the semantic and syntactic nuances of the tokens. Additionally, positional encoding is applied to each token to retain the order information, which is crucial in understanding the structure of the code.

The embedded tokens, now enriched with positional information, are subsequently processed by a stack of 12 transformer layers. These layers, through self-attention mechanisms, enable the model to capture dependencies between tokens regardless of their positions in the input sequence. The output from the final transformer layer corresponding to the \textless \textit{cls}\textgreater \ token, which aggregates the contextual information of the entire sequence, is then fed into a multi-class linear classification layer.

The classification layer produces logits, which are essentially raw predictions that have not yet been normalized. These logits are then transformed into probabilities using the softmax function, which assigns a probability to each class, indicating the likelihood of the input code belonging to a particular vulnerability class or being clean.

During the training phase, as illustrated in the bottom left of Fig.~\ref{fig:framework}, the output dimension is $n+1$, accounting for a `clean' class and $n$ vulnerable classes corresponding to the $n$ programming languages. This design mirrors our training data, which comprises labeled vulnerable functions from $n$ languages and their non-vulnerable counterparts. We employ the gradient descent algorithm to minimize the FOLA loss function, a variant designed to address the class imbalance issue, which we describe in Section \ref{sec:loss}. The model iteration with the highest performance on the validation set is selected as the final trained model.

In the testing phase, the model's output dimension is binary, distinguishing between `clean' and `vulnerable' classes. Here, the specific language of the vulnerability is not of interest; rather, the focus is on the binary determination of the presence of a vulnerability. To achieve this, the probabilities for the $n$ different vulnerable classes are summed up, resulting in a single probability representing the overall likelihood of the code being vulnerable. This aggregated probability is then compared to the probability of the `clean' class to make the final binary decision.

\subsection{Loss Function for Tackling Class Imbalance} \label{sec:loss}

In the domain of multi-class vulnerability detection, the class imbalance phenomenon presents a significant challenge~\cite{croft2022data}. Typically, the datasets used for training such models have a disproportionate number of examples across different classes. This imbalance often results in a model that is biased toward the majority class, leading to suboptimal performance on the minority classes, which are usually the more critical vulnerabilities to detect. We did not use sampling techniques to tackle class imbalance in our study as these techniques lose information in training data (i.e., under-sampling) and/or increase training size and time too significantly (i.e., over-sampling).

To mitigate this issue, we adopt a hybrid approach that combines the principles of Focal loss \cite{lin2017focal} with the logit adjustment method \cite{menon2021longtail}. The Focal loss function is designed to focus more on the hard-to-classify examples by reducing the relative loss for well-classified examples, thus putting more emphasis on correcting the misclassified examples. The logit adjustment method, on the other hand, aims to recalibrate the logits of each class based on their frequency, effectively adjusting the decision boundary for each class. The combined loss function, which we refer to as FOLA loss, is formulated as follows,
\begin{equation}
L_{FOLA} = -\alpha_t (1 - p_t)^\gamma \log(p_t) + \tau \log(q_t)
\end{equation}
where \( p_t \) is the model's estimated probability for the true class \( t \), \( \alpha_t \) is a weighting factor to balance the importance of different classes, and \( \gamma \) is the focusing parameter of the FOCAL loss that effectively reduces the loss contribution from easy examples and increases the importance of correcting misclassified examples. The term \( \tau \log(q_t) \) represents the logit adjustment for class \( t \), where \( \tau \) is a hyperparameter that controls the strength of the adjustment and \( q_t \) is the frequency of class \( t \).

By applying this FOLA loss function during the training of our MVD model, we can effectively address the class imbalance by dynamically adjusting the contribution of each class to the loss based on its frequency and the difficulty of classifying its examples. This ensures that the model does not become biased toward the majority class and improves its performance on the minority classes, which is crucial for achieving reliable performance of vulnerability detection across multiple programming languages.

\subsection{Extending to New Languages}
\label{subsec:incremental_learning_module}

In the realm of software development, the sheer number of programming languages presents a daunting challenge for vulnerability detection models. It is impractical to encompass all existing languages in the initial training phase of a model. In real-world applications, the deployed MVD model may encounter projects written in languages not included in its training corpus. Compounding this issue is the frequent scenario where users lack access to the original training data~\cite{nong2022open}, making it difficult to leverage past knowledge.

A naive solution might involve training a separate model for each new language from scratch, but this approach is fraught with inefficiencies, leading to model redundancy and a failure to capitalize on the knowledge embedded within the already trained MVD model.

To circumvent these issues, our MVD framework incorporates an incremental learning \cite{li2017learning} module. This module enables the model to extend its capabilities to new languages by building upon the knowledge acquired during its initial training. This method not only facilitates the learning of new tasks by leveraging existing capabilities but also largely maintains the model's performance in the original languages.

As shown in the Fig. \ref{fig:framework}, we implement this module by introducing a distillation loss that retains the knowledge of the original languages while accommodating new information. The equation for the distillation loss is as follows,
\begin{equation}
L_{distillation} = \sum_{i=1}^{N} (z_i - \sigma(z_i^{old}))^2
\end{equation}
where \( z_i \) represents the output logits of the new model for the original languages, \( \sigma \) denotes the softmax function, and \( z_i^{old} \) are the logits from the previously trained model. This loss ensures that the predictions for the original languages remain consistent before and after the model is updated.

The final loss function is a composite of the distillation loss and the FOLA loss, which addresses the class imbalance issue:
\begin{equation}
L_{total} = L_{distillation} + L_{FOLA}
\end{equation}

By optimizing this combined loss function during the training process, MVD can effectively learn to detect vulnerabilities in new programming languages while preserving its existing knowledge base. This approach streamlines the extension of the model's capabilities and ensures that the learned information is retained and utilized effectively.

\section{Experimental Design and Setup}
\label{sec:setup}

In this section, we describe the experimental design and setup for empirically evaluating our MVD framework.

\subsection{Research Questions}
\label{sec:rqs}

We set out to answer the following three Research Questions (RQs) to shed light on the effectiveness of MVD for multi-lingual software vulnerability detection.

\begin{itemize}
    \item \textbf{RQ1}: Can MVD outperform state-of-the-art models for software vulnerability detection in different languages?
    \item \textbf{RQ2}: What are the contributions of the key components in MVD to the model performance?
    \item \textbf{RQ3}: What is the performance of MVD when extended to a new language?
\end{itemize}

RQ1 seeks to evaluate the effectiveness of MVD against current leading models in the field of software vulnerability detection across various programming languages. Given MVD's architecture, which leverages the pre-training on data of multi-lingual vulnerabilities and a novel class-imbalance loss function, it is hypothesized that MVD can provide superior performance by effectively learning from a diverse set of vulnerability patterns across multiple languages. 
RQ2 aims to dissect the MVD framework to understand the impact of its individual components on the overall model performance. This involves analyzing the role of the multi-class classification paradigm, the FOLA loss function, and the strategy of fine-tuning either the entire model or just the classifier layer. By conducting an ablation study, we can determine how each component contributes to the model's ability to detect vulnerabilities and whether they are all critical to achieving the observed performance levels.
RQ3 addresses the model's adaptability and performance when extended to a new programming language not included in the initial training data. This question is crucial for understanding the practicality of MVD in real-world scenarios where it may need to be applied to languages that emerge or become relevant after the model has been deployed. RQ3  is expected to shed light on the extent to which the incremental learning module can integrate new languages without significant loss of performance on previously learned languages.

\subsection{Datasets}

We customized the methods and tools provided by CVEfixes~\cite{bhandari2021cvefixes} to curate vulnerability data for six different programming languages, namely C/C++, Python, Java, C\#, JavaScript, and TypeScript.
C and C++ were chosen because they have been commonly investigated in the literature (e.g.,~\cite{li2018vuldeepecker,li2021sysevr,fu2022linevul,hin2022linevd}). The other five are the most popular languages in practice, according to the developers' survey conducted by Stack Overflow.\footnote{\url{https://survey.stackoverflow.co/2023/\#most-popular-technologies-language-prof}} Note that we focused on general-purpose programming languages, so we excluded task-specific languages like SQL for database manipulations, HTML/CSS for web development, or Bash for scripting on Linux-based operating systems.
We first collected vulnerability-fixing commits in each of the aforementioned languages reported on the National Vulnerability Database~\cite{nvd_website}. In these commits, the functions encompassing lines changed were considered vulnerable; otherwise, they were non-vulnerable. Note that this data curation process follows the same practice of Big-Vul~\cite{fan2020ac}, the largest vulnerability dataset widely used in the literature.

To further increase the data quality, we applied a series of filtering steps to the collected functions. To ensure that a function was written in a particular language, we defined the file extensions for each language, as given in Table~\ref{tab:dataset}. Note that these extensions might not cover all available (vulnerable) code of each language in the wild, but they are the most commonly used ones in practice, ensuring the majority of code was curated. We also removed the functions inside test files to focus on production code. We also discarded functions that contained only cosmetic (non-functional) changes, e.g., changing whitespaces/newlines/comments as these functions were unlikely to contain vulnerabilities. These filtering steps are common practices in the literature (e.g.,~\cite{croft2021empirical,croft2022noisy,li2021vulnerability}).
We did not trace/include latent vulnerable functions as there is not yet an accurate way to automatically determine the origin (introduction time) of vulnerabilities~\cite{croft2022noisy}. After the filtering steps, the number of vulnerable and non-vulnerable functions are reported in Table~\ref{tab:dataset}. It is evident that the number of vulnerable functions was significantly smaller than that of non-vulnerable ones, confirming our argument about the existence of class imbalance in multi-lingual vulnerability prediction.

\begin{table}[!t]
\caption{The numbers of vulnerable and non-vulnerable functions along with the file extensions used for extracting the functions in each language.\label{tab:dataset}}
\centering
\begin{tabular}{ccccc}
\hline
\textbf{Language} & \textbf{Vuln.} & \textbf{Non-vuln.} & \textbf{\% Vuln.} & \textbf{Key file extension(s)}\\
\hline
Python & 779 & 10,801 & 6.7 & .py\\
C/C++ & 6,311 & 116,725 & 5.1 & .c, .cc, .cpp, .h, .hpp\\
Java & 789 & 10,687 & 6.9 & .java\\
C\# & 332 & 1,280 & 20.6 & .cs, .csx\\
JavaScript & 2,969 & 28,207 & 9.5 & .js, .jss\\
TypeScript & 151 & 1,760 & 7.9 & .ts, .tsx\\
\hline
\end{tabular}
\end{table}

\subsection{Evaluation Metrics}
In our experiments, we assessed the MVD framework using a variety of evaluation metrics that are widely accepted in vulnerability detection research. We incorporated the Area Under the Precision-Recall Curve (PR-AUC) as a key metric, which aggregates the precision-recall curve into a single value. PR-AUC is particularly advantageous in our setting as it evaluates the model's performance across all thresholds (threshold-agnostic), offering a measure that is unaffected by the selection of any specific decision boundary. This metric is also crucial for imbalanced classification like multi-lingual vulnerability detection (see Table~\ref{tab:dataset}), where the cost of missing a true vulnerability (low recall) and the expense of investigating a false alarm (low precision) must be carefully balanced. By using PR-AUC, we gain insight into the model's ability to discern between vulnerable and non-vulnerable code snippets across the entire spectrum of precision and recall, providing a robust indicator of its overall predictive quality.

We also employed the F1-score of the binary classification for its balanced consideration of precision and recall, making it particularly relevant for our imbalanced dataset where true negatives vastly outnumbered true positives. Precision is critical to ensure the model minimizes false positives, which can be costly and time-consuming in practical applications, while recall is essential for capturing as many true vulnerabilities as possible to maintain system security.
The Matthews correlation coefficient (MCC) was also used due to its effectiveness in providing a nuanced view of the model's performance across all quadrants of the confusion matrix, which is valuable in our context of imbalanced classes.

\subsection{Methodology for Answering RQ1}

In this research question, we aim to compare the performance of a model trained on a multi-lingual dataset encompassing all six languages against models trained exclusively on single-language datasets. For each language, we partitioned our dataset into training, validation, and testing sets following an 8:1:1 ratio, which has been the standard for vulnerability prediction (e.g.,~\cite{li2021vulnerability,fu2022linevul,hin2022linevd,steenhoek2023empirical}).
The model was trained on the training set, and its performance was assessed on the validation set after each epoch. The iteration achieving the highest PR-AUC on the validation set was preserved as the final model and tested on the testing set.

Regarding hyperparameters, we set the initial learning rate to \(2 \times 10^{-5}\) and employed a cosine annealing schedule, gradually reducing the learning rate in a cosine curve-like fashion as training progresses. This approach helps in fine-tuning the learning rate to converge optimally. We utilized the backpropagation algorithm and the AdamW optimizer \cite{loshchilov2018decoupled}, a variant of the Adam optimizer \cite{kingma2014adam} that is particularly effective for fine-tuning Transformer-based models. This optimizer updates the model weights to minimize the loss function.

Upon completion of the training phase, we evaluated the model's performance using the testing set to ensure an unbiased assessment of its generalization capabilities. For baseline comparisons, we adopted LineVul~\cite{fu2022linevul}, the state-of-the-art vulnerability prediction model~\cite{steenhoek2023empirical}, which involves fine-tuning CodeBERT using single-language vulnerability datasets. Consequently, we obtained distinct models for each language: LineVul-Python, LineVul-C/C++, LineVul-Java, LineVul-C\#, LineVul-JavaScript, and LineVul-TypeScript. We utilized the source code from LineVul\footnote{\url{https://github.com/awsm-research/LineVul}} and retrained the models using our datasets. The data split ratio, hyperparameters, model selection criteria, and evaluation procedures were consistent with those used for the multi-lingual model to ensure a fair comparison.

\subsection{Methodology for Answering RQ2}
To ascertain the individual impact of MVD's components, we conducted an ablation study using the same evaluation setup as in RQ1. We systematically removed or altered certain components to observe the change in performance, thereby validating the significance of each component.

Firstly, we compared the full MVD model, which employs a multi-class classification paradigm, with a variant we term MVD-binary. The MVD-binary model simplifies the problem by aggregating all vulnerable examples into a single class, regardless of the language. This binary classification approach aligns with traditional single-language vulnerability detection models, where the classifier is binary. By comparing the performance of MVD-binary with the full MVD model, we can assess the efficacy of the multi-class approach in enhancing the model's discriminative power across multiple languages.

Next, we turned our attention to the FOLA loss function, which is a composite of Focal loss and logit adjustment. To evaluate its effectiveness, we trained variants of the MVD model using different loss functions: one with Focal loss alone, one with cross-entropy combined with logit adjustment, and one with the standard cross-entropy loss. By comparing these variants, we can determine the contribution of the FOLA loss function to the model's ability to handle class imbalance and improve performance for minority classes.

Lastly, we explored the utility of using the base CodeBERT model solely as a feature extractor, wherein its weights remain frozen during the training of the classifier. This approach can preserve the original representations learned by CodeBERT and expedite the training process. By comparing this method with the full model training, where CodeBERT's weights are fine-tuned during training, we can discern whether the additional fine-tuning step significantly contributes to the model's performance or if the pre-trained representations are sufficient for vulnerability detection tasks.

Through this ablation study, we aim to shed light on the necessity and efficiency of each component and training strategy within the MVD framework, providing insights into their roles in achieving the model's overall performance.

\subsection{Methodology for Answering RQ3}
Our experiment was designed to investigate the adaptability of the MVD model when it is extended to accommodate a new programming language. This process was conducted in two distinct stages to simulate the scenario where a previously unencountered language needs to be integrated into an existing model. The data splits were the same as in RQ1.

Initially, we prepared the groundwork by training six separate MVD models, each intentionally omitting one of the languages from the training data. This language, excluded in the first stage, was designated as the `new' language for the subsequent phase of the experiment. By doing so, we created a baseline for how the model performs without any prior knowledge of the new language.

In the second stage, we employed the incremental learning module, described in Section~\ref{subsec:incremental_learning_module}, to introduce the new language to the pre-trained MVD models. This step allows us to observe how the model assimilates new information and whether it can leverage the knowledge acquired from the original languages to enhance its performance on the new language.

Upon completion of the incremental learning process, we conducted a series of comparisons to evaluate the efficacy of this approach. We measured the performance of the MVD model on the new language and compared it with that of a single-language vulnerability detection model trained solely on the new language. This comparison aims to highlight the advantages of using a multi-lingual model that can transfer learned knowledge to new contexts/languages.

Furthermore, we assessed the performance of the original languages both before and after the application of incremental learning. This comparison is essential to ensure that the extension process does not detrimentally affect the model's existing capabilities.

Finally, we compared the performance of the incrementally updated model with the MVD model that was trained with all six languages from the outset. This comparison is intended to illustrate the gap, if any, between the incrementally learned model and the theoretical optimum, where the model has been trained on all languages simultaneously.

Overall, RQ3 aims to not only validate the incremental learning approach but also to quantify its impact on both the new and original languages, thereby providing a better understanding of the model's extensibility and robustness in the face of evolving software development practices.

\section{Experimental Results}
\label{sec:results}
We present the experimental results of our proposed model, MVD, per the methods described in Section~\ref{sec:setup}.

\subsection{RQ1: MVD vs. Single-Language Baselines}
\label{subsec:rq1_results}

Table~\ref{tab:rq1} presents a comparative analysis between the performance of the multi-lingual Vulnerability Detection (MVD) model and the single-language LineVul models based on CodeBERT, which were trained on individual programming languages. Each sub-table is titled with the language used for testing, and the rows labeled LineVul-\textit{language} represent the LineVul models trained specifically for that \textit{language}. The colors red and blue in the table highlight the top-1 and top-2 performance metrics, respectively.

The experimental results showcased that the MVD model consistently achieved top-tier performance, either ranking first or at least second, often outperforming the single-language LineVul models trained on their respective languages.
The significant improvements included 34.9\% for C/C++, 30.7\% for Java, and 24.9\% for TypeScript in terms of PR-AUC.
For the remaining languages, MVD performed on par (within 5\% in PR-AUC) compared to that of the single-language LineVul counterparts.
The general trend of the MVD model outperforming the baselines was also observed for the other metrics.
These results confirm the effectiveness of our unified MVD model, confirming that training across multiple languages can leverage cross-linguistic knowledge of vulnerabilities and significantly improve vulnerability detection efficacy.

Overall, the results illustrated that a single MVD model could effectively operate across different languages, unlike the LineVul models, which often exhibited substantial performance declines when tested outside their training language.
On average, MVD had 83.7\%, 167.2\%, 137.9\%, 101.5\%, 125.6\%, and 193.6\% better performance (PR-AUC) of predicting vulnerabilities in all six languages than the state-of-the-art LineVul models trained specifically for Python, C/C++, Java, and JavaScript, C\# and TypeScript, respectively. It is also worth noting that MVD was approximately 7\% better PR-AUC than that (0.5008) of the LineVul models trained for each language individually and requiring nearly five times more resources.
All these results highlight the MVD model's superior capability to identify vulnerabilities in software projects developed in multiple languages, thereby enhancing its practical utility in diverse development environments.

\begin{table}[t]
\centering
\caption{The comparison between our MVD model and the baseline single language vulnerability detection models. \textbf{Note}: For a given language, the red and blue colors denote the top-1 and top-2 values of each metric for that language.}
\begin{tabular}{lccccc}
\hline
\textbf{Model} & \textbf{PR-AUC} & \textbf{F1} & \textbf{Precision} & \textbf{Recall} & \textbf{MCC}  \\ 
\hline
\multicolumn{6}{c}{\textbf{Python}} \\
\hline

LineVul-Python & \textcolor{blue}{0.8824} & \textcolor{blue}{0.8810} & \textcolor{blue}{0.8101} & \textcolor{red}{\textbf{0.9669}} & \textcolor{blue}{0.8777}  \\
LineVul-C/C++ & 0.1330 & 0.0022 & 0.0333 & 0.0011 & 0.0029  \\
LineVul-Java & 0.1360 & 0.0927 & 0.1569 & 0.0730 & 0.0623  \\
LineVul-JavaScript & 0.1007 & 0.0591 & 0.0840 & 0.0527 & 0.0204  \\
LineVul-C\# & 0.0948 & 0.1015 & 0.1187 & 0.1503 & 0.0567  \\
LineVul-TypeScript & 0.1316 & 0.0665 & 0.1224 & 0.0567 & 0.0471  \\
\textbf{MVD} & \textcolor{red}{\textbf{0.8875}} & \textcolor{red}{\textbf{0.8830}} & \textcolor{red}{\textbf{0.9731}} & \textcolor{blue}{0.8098} & \textcolor{red}{\textbf{0.8804}} \\
\hline
\multicolumn{6}{c}{\textbf{C/C++}} \\
\hline
LineVul-Python & 0.0908 & 0.1082 & 0.0897 & 0.1368 & 0.0518 \\
LineVul-C/C++ & \textcolor{blue}{0.2534} & \textcolor{blue}{0.1458} & \textcolor{blue}{0.7086} & 0.0825 & \textcolor{blue}{0.2266} \\
LineVul-Java & 0.1201 & 0.0282 & 0.3325 & 0.0150 & 0.0564 \\
LineVul-JavaScript & 0.1059 & 0.0201 & 0.4209 & 0.0104 & 0.0554 \\
LineVul-C\# & 0.0771 & 0.1182 & 0.0845 & \textcolor{red}{\textbf{0.2743}} & 0.0664 \\
LineVul-TypeScript & 0.0809 & 0.0940 & 0.0926 & 0.2207 & 0.0517 \\
\textbf{MVD} & \textcolor{red}{\textbf{0.3418}} & \textcolor{red}{\textbf{0.2255}} & \textcolor{red}{\textbf{0.7695}} & \textcolor{blue}{0.1345} & \textcolor{red}{\textbf{0.3048}}  \\
\hline
\multicolumn{6}{c}{\textbf{Java}} \\
\hline
LineVul-Python & 0.1277 & 0.1004 & 0.1411 & 0.0787 & 0.0559 \\
LineVul-C/C++ & 0.1911 & 0.0710 & \textcolor{red}{\textbf{0.6528}} & 0.0378 & 0.1482 \\
LineVul-Java & \textcolor{blue}{0.3216} & \textcolor{blue}{0.2688} & 0.2907 & \textcolor{blue}{0.2500} & \textcolor{blue}{0.2190}  \\
LineVul-JavaScript & 0.1860 & 0.1196 & \textcolor{blue}{0.6283} & 0.0677 & 0.1844 \\
LineVul-C\# & 0.0844 & 0.0335 & 0.0571 & 0.0292 & 0.0074 \\
LineVul-TypeScript & 0.1452 & 0.0072 & 0.1367 & 0.0039 & 0.0140 \\
\textbf{MVD} & \textcolor{red}{\textbf{0.4204}} & \textcolor{red}{\textbf{0.3317}} & 0.5693 & \textcolor{red}{\textbf{0.2584}} & \textcolor{red}{\textbf{0.3368}} \\
\hline
\multicolumn{6}{c}{\textbf{C\#}} \\
\hline
LineVul-Python & 0.2510 & 0.0542 & 0.1492 & 0.0345 & -0.0316 \\
LineVul-C/C++ & 0.3471 & 0.0655 & 0.6000 & 0.0349 & 0.1287 \\
LineVul-Java & 0.3432 & 0.1094 & 0.4923 & 0.0658 & 0.1254 \\
LineVul-JavaScript & 0.3636 & 0.1244 & \textcolor{red}{\textbf{0.8833}} & 0.0684 & 0.2040 \\
LineVul-C\# & \textcolor{red}{\textbf{0.7427}} & \textcolor{red}{\textbf{0.6582}} & 0.7685 & \textcolor{red}{\textbf{0.5845}} & \textcolor{red}{\textbf{0.5990}} \\
LineVul-TypeScript & 0.3214 & 0.0394 & 0.3050 & 0.0211 & 0.0640 \\
\textbf{MVD} & \textcolor{blue}{0.7352} & \textcolor{blue}{0.6475} & \textcolor{blue}{0.7866} & \textcolor{blue}{0.5700} & \textcolor{blue}{0.5948} \\
\hline
\multicolumn{6}{c}{\textbf{JavaScript}} \\
\hline
LineVul-Python & 0.1482 & 0.0958 & 0.1944 & 0.0638 & 0.0598 \\
LineVul-C/C++ & 0.2030 & 0.0554 & 0.6173 & 0.0292 & 0.1225 \\
LineVul-Java & 0.1888 & 0.1701 & 0.2328 & 0.1773 & 0.1131 \\
LineVul-JavaScript & \textcolor{red}{\textbf{0.5594}} & \textcolor{red}{\textbf{0.4625}} & \textcolor{blue}{0.5272} & \textcolor{blue}{0.4119} & \textcolor{red}{\textbf{0.4767}} \\
LineVul-C\# & 0.1341 & 0.1960 & 0.1275 & \textcolor{red}{\textbf{0.4960}} & 0.0746 \\
LineVul-TypeScript & 0.1416 & 0.1278 & 0.1565 & 0.1571 & 0.0599 \\
\textbf{MVD} & \textcolor{blue}{0.5345} & \textcolor{blue}{0.4224} & \textcolor{red}{\textbf{0.6858}} & 0.3136 & \textcolor{blue}{0.4233} \\
\hline
\multicolumn{6}{c}{\textbf{TypeScript}} \\
\hline
LineVul-Python & 0.1556 & 0.1006 & 0.1595 & 0.0770 & 0.0540 \\
LineVul-C/C++ & 0.2801 & 0.0826 & 0.4000 & 0.0476 & 0.1271 \\
LineVul-Java & 0.2241 & 0.1517 & 0.2921 & 0.1124 & 0.1265 \\
LineVul-JavaScript & \textcolor{blue}{0.2454} & \textcolor{blue}{0.1753} & \textcolor{blue}{0.4847} & 0.1159 & \textcolor{blue}{0.1961} \\
LineVul-C\# & 0.1569 & 0.1504 & 0.1298 & \textcolor{red}{\textbf{0.2222}} & 0.0746 \\
LineVul-TypeScript & 0.1234 & 0.0769 & 0.0667 & 0.0909 & 0.0248 \\
\textbf{MVD} & \textcolor{red}{\textbf{0.3065}} & \textcolor{red}{\textbf{0.1833}} & \textcolor{red}{\textbf{0.5690}} & \textcolor{blue}{0.1167} & \textcolor{red}{\textbf{0.2265}} \\
\hline
\multicolumn{6}{c}{\textbf{Average}} \\
\hline
LineVul-Python & \textcolor{blue}{0.2926} & 0.2234 & 0.2573 & 0.2263 & 0.1656 \\ 
LineVul-C/C++ & 0.2012 & 0.0799 & \textcolor{blue}{0.4187} & 0.0389 & 0.0982 \\ 
LineVul-Java & 0.2260 & 0.1047 & 0.2805 & 0.0982 & 0.0958 \\ 
LineVul-JavaScript & 0.2602 & 0.1601 & 0.5047 & 0.1215 & \textcolor{blue}{0.1728} \\ 
LineVul-C\# & 0.2383 & \textcolor{blue}{0.2266} & 0.2215 & \textcolor{blue}{0.2911} & 0.1442 \\ 
LineVul-TypeScript & 0.1831 & 0.0674 & 0.1406 & 0.0752 & 0.0420 \\ 
\textbf{MVD} & \textcolor{red}{\textbf{0.5376}} & \textcolor{red}{\textbf{0.4489}} & \textcolor{red}{\textbf{0.7255}} & \textcolor{red}{\textbf{0.3672}} & \textcolor{red}{\textbf{0.4611}} \\ 

\hline
\end{tabular}
\label{tab:rq1}
\end{table}

\subsection{RQ2: Ablation Study of MVD's Components}
\label{subsec:rq2_results}

The impacts of the components on the performance of our MVD model are shown in Table \ref{tab:rq2}. The different variants of the MVD model included in the ablation study are described and analyzed hereafter.

Firstly, we compared the full MVD model, which employs a multi-class classification paradigm, with a variant termed \textit{MVD-binary}. The MVD-binary model simplifies the problem by aggregating all vulnerable examples into a single class, irrespective of the language. This binary classification approach aligns with traditional single-language vulnerability detection models, where the classifier is binary. While the MVD-binary model performed competitively, the full MVD model, utilizing a multi-class approach, outperformed the MVD-binary model in all metrics across all languages (only except Recall in JavaScript), as well as by 4\% in PR-AUC, on average. This highlights the improved efficacy of the multi-class approach over the conventional binary counterpart.

We next assessed the efficacy of the FOLA loss function, a hybrid of Focal loss and logit adjustment. Variants of the MVD model were trained using distinct loss functions: \textit{MVD-focal}, employing Focal loss; \textit{MVD-lace}, combining cross-entropy with logit adjustment; and \textit{MVD-ce}, using standard cross-entropy. The outcomes demonstrated that the FOLA loss function was superior in managing class imbalance and enhancing performance in minority classes, consistently achieving at least top-2 PR-AUC across all languages. Conversely, models employing other loss functions exhibited performance variability across different languages, complicating the task of achieving balanced performance. Regarding the average performance, the MVD outperformed the models with all other loss functions by at least 0.2\% in PR-AUC. These results substantiate the beneficial impact of the FOLA loss function.

Lastly, we explored the utility of using the base CodeBERT model solely as a feature extractor (MVD-freeze), with its weights remaining frozen during the training of the classifier. This approach aimed to preserve the original representations learned by CodeBERT and expedite the training process. However, the results indicated that MVD-freeze struggled to achieve comparable performance to the fully fine-tuned MVD. Specifically, the average performance showed a significant gap of around 30\% in PR-AUC. These findings suggest that fine-tuning CodeBERT during training significantly enhances the model's performance.

All the above observations elucidate the necessity and efficiency of every component and training strategy within the MVD framework. The results have provided insights into the roles of multi-class classification, the FOLA loss function, and fine-tuning of pre-trained models in achieving superior performance in multi-lingual vulnerability detection, confirming the overall effectiveness of our MVD framework.

\begin{table}[t]
\centering
\caption{The comparison between our MVD model and its variants for ablation analysis. \textbf{Note}: For a given language, the red and blue colors denote the top-1 and top-2 values of each metric for that language.}
\begin{tabular}{lccccc}
\hline
\textbf{Model} & \textbf{PR-AUC} & \textbf{F1} & \textbf{Precision} & \textbf{Recall} & \textbf{MCC}  \\ 
\hline
\multicolumn{6}{c}{\textbf{Python}} \\
\hline

MVD-binary & 0.8750 & 0.8757 & 0.9618 & 0.8047 & 0.8721  \\
MVD-ce & 0.8821 & \textcolor{blue}{0.8848} & \textcolor{red}{\textbf{0.9842}} & 0.8051 & \textcolor{blue}{0.8832} \\
MVD-focal & \textcolor{red}{\textbf{0.8886}} & \textcolor{red}{\textbf{0.8867}} & \textcolor{blue}{0.9822} & 0.8093 & \textcolor{red}{\textbf{0.8848}}  \\
MVD-lace & 0.8810 & 0.8847 & 0.9757 & \textcolor{red}{\textbf{0.8105}} & 0.8822  \\
MVD-freeze & 0.4249 & 0.2229 & 0.7482 & 0.1326 & 0.2988 \\
\textbf{MVD} & \textcolor{blue}{0.8875} & 0.8830 & 0.9731 & \textcolor{blue}{0.8098} & 0.8804  \\
\hline
\multicolumn{6}{c}{\textbf{C/C++}} \\
\hline
MVD-binary & 0.2986 & \textcolor{red}{\textbf{0.2519}} & 0.5352 & \textcolor{red}{\textbf{0.1768}} & 0.2755 \\
MVD-ce & 0.3387 & 0.2219 & \textcolor{blue}{0.7946} & 0.1304 & \textcolor{blue}{0.3076} \\
MVD-focal & \textcolor{red}{\textbf{0.3456}} & \textcolor{blue}{0.2260} & \textcolor{red}{\textbf{0.7958}} & \textcolor{blue}{0.1348} & \textcolor{red}{\textbf{0.3098}} \\
MVD-lace & 0.3295 & 0.2204 & 0.7442 & 0.1312 & 0.2966 \\
MVD-freeze & 0.1326 & 0 & 0 & 0 & 0 \\
\textbf{MVD} & \textcolor{blue}{0.3418} & 0.2255 & 0.7695 & 0.1345 & 0.3048 \\
\hline
\multicolumn{6}{c}{\textbf{Java}} \\
\hline
MVD-binary & 0.3626 & 0.2927 & 0.5335 & \textcolor{blue}{0.2215} & 0.2981 \\
MVD-ce & 0.4144 & 0.3197 & 0.7342 & 0.2095 & \textcolor{blue}{0.3657}  \\
MVD-focal & \textcolor{blue}{0.4193} & \textcolor{blue}{0.3123} & \textcolor{blue}{0.7344} & 0.2130 & 0.3586 \\
MVD-lace & 0.4125 & 0.3066 & \textcolor{red}{\textbf{0.7830}} & 0.1962 & \textcolor{red}{\textbf{0.3660}}\\
MVD-freeze & 0.1591 & 0.0625 & 0.2214 & 0.0368 & 0.0706 \\
\textbf{MVD} & \textcolor{red}{\textbf{0.4204}} & \textcolor{red}{\textbf{0.3317}} & 0.5693 & \textcolor{red}{\textbf{0.2584}} & 0.3368 \\
\hline
\multicolumn{6}{c}{\textbf{C\#}} \\
\hline
MVD-binary & 0.6945 & \textcolor{blue}{0.6036} & 0.7759 & 0.5140 & 0.5555 \\
MVD-ce & \textcolor{red}{\textbf{0.7443}} & 0.6016 & \textcolor{blue}{0.7930} & 0.4989 & 0.5570 \\
MVD-focal & 0.7294 & 0.5919 & 0.7343 & \textcolor{blue}{0.5310} & 0.5325 \\
MVD-lace & 0.7205 & 0.6008 & \textcolor{red}{\textbf{0.8540}} & 0.4724 & \textcolor{blue}{0.5721} \\
MVD-freeze & 0.3064 & 0 & 0 & 0 & -0.0094 \\
\textbf{MVD} & \textcolor{blue}{0.7352} & \textcolor{red}{\textbf{0.6475}} & 0.7866 & \textcolor{red}{\textbf{0.5700}} & \textcolor{red}{\textbf{0.5948}} \\
\hline
\multicolumn{6}{c}{\textbf{JavaScript}} \\
\hline
MVD-binary & 0.4912 & \textcolor{blue}{0.4209} & 0.5986 & \textcolor{red}{\textbf{0.3473}} & 0.4022 \\
MVD-ce & \textcolor{blue}{0.5315} & 0.4013 & \textcolor{red}{\textbf{0.7125}} & 0.2818 & \textcolor{blue}{0.4139} \\
MVD-focal & 0.5226 & 0.3986 & \textcolor{blue}{0.6957} & 0.2866 & 0.4072  \\
MVD-lace & 0.5100 & 0.4029 & 0.6576 & 0.296 & 0.4015 \\
MVD-freeze & 0.2196 & 0.0032 & 0.1556 & 0.0016 & 0.0104 \\
\textbf{MVD} & \textcolor{red}{\textbf{0.5345}} & \textcolor{red}{\textbf{0.4224}} & 0.6858 & \textcolor{blue}{0.3136} & \textcolor{red}{\textbf{0.4233}} \\
\hline
\multicolumn{6}{c}{\textbf{TypeScript}} \\
\hline
MVD-binary & 0.2663 & \textcolor{blue}{0.1412} & 0.3482 & \textcolor{blue}{0.1032} & 0.1384 \\
MVD-ce & 0.3000 & 0.1155 & \textcolor{blue}{0.5333} & 0.0709 & \textcolor{blue}{0.1635} \\
MVD-focal & 0.2899 & 0.1335 & 0.4067 & 0.0818 & 0.1605 \\
MVD-lace & \textcolor{red}{\textbf{0.3150}} & 0.1294 & 0.4231 & 0.0801 & 0.1610 \\
MVD-freeze & 0.2129 & 0 & 0 & 0 & 0 \\
\textbf{MVD} & \textcolor{blue}{0.3065} & \textcolor{red}{\textbf{0.1833}} & \textcolor{red}{\textbf{0.5690}} & \textcolor{red}{\textbf{0.1167}} & \textcolor{red}{\textbf{0.2265}} \\
\hline
\multicolumn{6}{c}{\textbf{Average}} \\
\hline
MVD-binary & 0.4980 & \textcolor{blue}{0.4310} & 0.5422 & \textcolor{blue}{0.3612} & 0.4236 \\
MVD-ce & \textcolor{blue}{0.5351} & 0.4241 & \textcolor{blue}{0.6253} & 0.3328 & \textcolor{blue}{0.4485} \\
MVD-focal & 0.5326 & 0.4215 & 0.6025 & 0.3428 & 0.4423 \\
MVD-lace & 0.5281 & 0.4235 & 0.5729 & 0.3310 & 0.4466 \\
MVD-freeze & 0.2426 & 0.0481 & 0.1875 & 0.0285 & 0.0617 \\
\textbf{MVD} & \textcolor{red}{\textbf{0.5376}} & \textcolor{red}{\textbf{0.4489}} & \textcolor{red}{\textbf{0.7255}} & \textcolor{red}{\textbf{0.3672}} & \textcolor{red}{\textbf{0.4611}} \\
\hline
\end{tabular}
\label{tab:rq2}
\end{table}

\subsection{RQ3: Extension of MVD to New Languages}
\label{subsec:rq3_results}
This experiment was conducted to assess the adaptability of the MVD model when a new programming language is integrated incrementally and the training data of old language(s) is no longer accessible. The results in Table \ref{tab:rq3} highlighted several promises of the effectiveness of incremental learning compared to training a model from scratch with all languages.

The incremental learning approach (inc-X, where X represents the newly introduced language) was generally better for four out of six languages in PR-AUC for the new language compared to models that were trained on that language alone (LineVul). This trend indicates that the incremental learning approach can effectively assimilate new information and improve the model's performance in the newly added language.

The results also revealed that the performance on the original languages did not degrade significantly and could even improve following the incremental learning process. For example, when C/C++ was incrementally added, the performance of the model on Python decreased from 0.9018 (w/o-C/C++) to 0.8846 (inc-C/C++), which suggests some loss. However, we also witnessed performance increase after incremental learning; for instance, when Java was incrementally added, the performance on TypeScript even improved (from 0.1653 to 0.2860).
This suggests that the model retains much of its original knowledge even when the training data of old languages are unavailable.
Further, we observed that the performance of MVD for each of the six languages after incrementally extending to a new language could vary. For example, the PR-AUC of MVD for TS ranged from 0.1260 to 0.2860 when incrementally learned in different languages. In addition, incrementally training MVD in a new language did not always lead to better performance for that language than incremental training in other languages. These results imply that the language-wise performance of MVD after incremental learning depends on the combination and order of the languages on which the model was previously trained.

\begin{table}[t]
\centering
\caption{The comparison in terms of PR-AUC when expanding to new languages with incremental learning. \textbf{Note}: For a given extension to a new language, the red color denotes the best value of each language for that scenario.}
\begin{tabular}{lcccccc}
\hline
\textbf{Model} & \textbf{Python} & \textbf{C/C++} & \textbf{Java} & \textbf{JS} & \textbf{C\#} & \textbf{TS} \\ 
\hline

LV-Python & 0.8824 & - & - & - & - & -\\
w/o-Python & - & \textcolor{red}{\textbf{0.2863}} & \textcolor{red}{\textbf{0.3827}} & \textcolor{red}{\textbf{0.4787}} & \textcolor{black}{0.7088} & 0.1438 \\
inc-Python & \textcolor{red}{\textbf{0.8920}} & 0.2691 & 0.3427 & 0.4665 & \textcolor{red}{\textbf{0.7265}} & \textcolor{red}{\textbf{0.1914}}\\
\hline
LV-C/C++ & - & 0.2534 & - & - & - & - \\
w/o-C/C++ & \textcolor{red}{\textbf{0.9018}} & - & 0.3541 & \textcolor{red}{\textbf{0.4862}} & 0.6877 & 0.0901 \\
inc-C/C++ & 0.8846 & \textcolor{red}{\textbf{0.3241}} & \textcolor{red}{\textbf{0.3777}} & 0.4117 & \textcolor{red}{\textbf{0.7069}} & \textcolor{red}{\textbf{0.1260}} \\
\hline
LV-Java & - & - & 0.3216 & - & - & - \\
w/o-Java & \textcolor{red}{\textbf{0.8898}} & \textcolor{red}{\textbf{0.2951}} & - & \textcolor{red}{\textbf{0.5146}} & 0.6403 & 0.1653 \\
inc-Java & 0.8059 & 0.2915 & \textcolor{red}{\textbf{0.3398}} & 0.4561 & \textcolor{red}{\textbf{0.6932}} & \textcolor{red}{\textbf{0.2860}} \\
\hline
LV-JS & - & - & - & \textcolor{red}{\textbf{0.5594}} & - & -\\
w/o-JS & \textcolor{red}{\textbf{0.8924}} & \textcolor{red}{\textbf{0.2969}} & 0.3316 & - & 0.6529 & 0.0886 \\
inc-JS & 0.8873 &  0.2897 &  \textcolor{red}{\textbf{0.3745}} &   0.4361 & \textcolor{red}{\textbf{0.6886}} & \textcolor{red}{\textbf{0.1960}} \\
\hline
LV-C\# & - & - & - & - & \textcolor{red}{\textbf{0.7427}} & - \\
w/o-C\# & \textcolor{red}{\textbf{0.8966}} & 0.3214 & \textcolor{red}{\textbf{0.3777}} & 0.4428 & - & \textcolor{red}{\textbf{0.2569}} \\
inc-C\# & 0.8964 & \textcolor{red}{\textbf{0.3216}} &  0.3500 &  \textcolor{red}{\textbf{0.5052}} & 0.7081 &  0.2407 \\
\hline

LV-TS & - & - & - & - & - & 0.1234 \\
w/o-TS & \textcolor{red}{\textbf{0.8906}} & 0.2978 & 0.3519 & 0.4831 & 0.6454 & - \\
inc-TS &  0.8899 & \textcolor{red}{\textbf{0.3140}} &  \textcolor{red}{\textbf{0.3726}} &  \textcolor{red}{\textbf{0.5027}} &  \textcolor{red}{\textbf{0.6949}} &  \textcolor{red}{\textbf{0.1531}} \\
\hline
MVD & 0.8875 & 0.3418 & 0.4204 & 0.5345 & 0.7352 & 0.3065 \\
\hline
\end{tabular}
\label{tab:rq3}
\end{table}

Furthermore, we compared the incrementally trained models (denoted as inc-X) with the MVD model trained on all six languages. While the incrementally trained models could occasionally outperform the full MVD model on specific languages (e.g., inc-Python slightly outperforming the MVD model for Python by 0.5\% in PR-AUC), the full MVD model generally achieved higher PR-AUC across all languages. This suggests that the MVD model benefits from simultaneous multi-lingual training, capturing diverse and shared patterns and features that lead to a more stable and balanced performance overall. However, incremental learning remains an effective strategy for efficiently adapting the model to new languages without retraining from scratch, despite not always reaching the peak performance of a fully trained multi-lingual model using the data of all available languages.

Overall, the experimental results have demonstrated the efficacy of the incremental learning approach in extending the MVD model to accommodate new languages while largely preserving its performance on previously known languages. Although the incrementally trained models do not achieve the same level of performance as a model trained with all languages from the start, they still provide a viable solution for environments where retraining on all data is impractical.

\section{Threats to Validity}\label{sec:threats_to_validity}

\subsection{Threats to Construct Validity}
The threats to construct validity concern the data selection in multiple programming languages. We utilized the methods and tools provided by CVEfixes, one of the largest multi-lingual vulnerability datasets in the literature, for our data collection. This dataset follows the latest practice to curate vulnerable and non-vulnerable functions. On top of the data provided by CVEfixes, we also improved the quality by removing the code irrelevant to vulnerability based on the recent checklist of vulnerability data quality assessment~\cite{croft2023data}.

\subsection{Threats to Internal Validity}
The internal validity threats are related to the optimality of the vulnerability prediction models. With limited computational resources, we could not try all possible hyperparameters. We still tuned our models using the common hyperparameters from relevant studies. For the LineVul baseline model, we also leveraged the recommended hyperparameters to tune it. The performance of our proposed MVD model may not be the highest possible for multi-lingual vulnerability prediction, but at least it established a strong foundation for future work to compare with and build upon.

\subsection{Threats to External Validity}
The external validity threats are pertinent to the generalizability of our findings.
We performed experiments and analysis of MVD using data from hundreds of projects in six different languages and various application domains, but our findings may still not generalize to other languages.

\section{Conclusion and Future Work}
\label{sec:conclusions}

We introduce MVD, a novel framework for multi-lingual software vulnerability detection that addresses the limitations of existing single-language-focused approaches.
Specifically, MVD is a unified model that is capable of predicting the existence of vulnerable functions written in multiple languages at the same time.
By leveraging a curated dataset of over 11,000 real-world vulnerabilities across six popular programming languages, MVD demonstrates superior detection performance by 83.7\% to 193.6\% compared to state-of-the-art models.
Our novel use of incremental learning also enables seamless extension to new languages without significantly degrading performance on previously supported ones, even in the absence of prior training data.
The promising results of MVD are envisioned to inspire future research into innovative approaches for managing vulnerabilities in modern multi-lingual software ecosystems. To advance toward this vision, we plan to enhance MVD by incorporating support for additional programming languages and extending its capabilities to predict crucial information such as exploitability, impact, and severity following the detection step. These enhancements aim to empower developers with deeper insights to effectively understand and address detected vulnerabilities.

\section{Data Availability}
\label{sec:data_avail}
The data and code of this study are available at~\url{https://figshare.com/s/10ec70108294a225f391}.

\section*{Acknowledgments}
The work was supported by the Cyber Security Research Centre Limited whose activities are partially funded by the Australian Government's Cooperative Research Centres Programme. This work was supported with supercomputing resources provided by the Phoenix HPC service at the University of Adelaide.

\IEEEtriggeratref{46}
\bibliographystyle{IEEEtran}
\bibliography{reference}

\end{document}